\documentclass[useAMS,usenatbib,usegraphicx]{mn2e}
\usepackage{epsf}
\usepackage{amssymb,amsmath}

\title[Measuring the viscosity of dark matter]{Measuring the viscosity of dark matter with strongly lensed gravitational waves}

\author[Cao et al.]
{Shuo Cao$^{1\ast}$, Jingzhao Qi$^2$, Marek Biesiada$^{1,3\dag}$, Tonghua Liu$^1$, Jin Li$^{4}$, Zong-Hong Zhu$^{1\ddag}$  \\
$^1$ Department of Astronomy, Beijing Normal University, 100875, Beijing, China; \emph{caoshuo@bnu.edu.cn; zhuzh@bnu.edu.cn}\\
$^2$ Department of Physics, College of Sciences, Northeastern University, Shenyang 110004, China;\\
$^3$ National Centre for Nuclear Research, Pasteura 7, 02-093 Warsaw, Poland; \emph{Marek.Biesiada@ncbj.gov.pl}\\
$^4$ Department of physics, Chongqing University, 400044 Chongqing,
China}

\begin{document}

\date{\today}

\voffset- .5in

\pagerange{\pageref{firstpage}--\pageref{lastpage}} \pubyear{}

\maketitle

\label{firstpage}

\begin{abstract}

Based on the strongly lensed gravitational waves (GWs) from compact
binary coalescence, we propose a new strategy to examine the fluid
shear viscosity of dark matter (DM) in the gravitational wave
domain, i.e., whether a GW experiences the damping effect when it
propagates in DM fluid with nonzero shear viscosity. By assuming
that the dark matter self-scatterings are efficient enough for the
hydrodynamic description to be valid, our results demonstrate that
future ground-based Einstein Telescope (ET) and satellite GW
observatory (Big Bang Observer; BBO) may succeed in detecting any
dark matter self-interactions at the scales of galaxies and
clusters.

\end{abstract}


%%%%%%%%%%%%%%%%%%%%%%%%%%%%%%%%%%%%%%%%%%%%%%%%%%%%%%%%%%%%%%%%%%%%%%%%%%%%%%
%%%%%%%%%%%%%%%%%%%%%%%%%%%%%%%%%%%%%%%%%%%%%%%%%%%%%%%%%%%%%%%%%%%%%%%%%%%%%%

\section{Introduction}

It is well known that the past years have witnessed a new era of
gravitational wave astronomy, with the first direct detection of
gravitational wave (GW) sources from inspiraling and merging binary
black holes (BH) \citep{Abbott16a} and neutron stars (NS)
\citep{Abbott17c}. These GW signals, together with their
electromagnetic (EM) counterparts can be used as standard sirens
providing the luminosity distances ($D_L$) in a direct way, not
relying on the cosmic distance ladder \citep{Schutz86}. Currently,
the measurements of luminosity distance by the advanced LIGO and
Virgo detectors are affected by large uncertainties. However, as
discussed in \citet{Gupta2019} the accuracy of the third generation
of GW detectors will reach $1-3\%$ accuracy up to 300 Mpc, much
smaller than velocity of host galaxies. This would allow for
independent calibration of distances to SNe Ia. Moreover, extending
the GW frequency window by space-borne detectors to decihertz band,
where the coalescences of binary white dwarfs are detectable, would
allow to measure luminosity distance up to 100 Mpc with $1\%$
accuracy \citep{Maselli2020,Zou2020}. Therefore, multi-messenger
observations can put strong constraints on the Hubble constant
inferred from low redshift sources. There also exist some
interesting possibilities of testing the speed of GWs by measuring
the time of arrival delays between photons and GWs over cosmological
distances~\citep{Nishizawa2016,Li2016}, the equivalence principle by
using the Shapiro effect~\citep{Will2014,Kahya2016}, and the Lorentz
invariance violation~\citep{Kostelecky2016}.

From theoretical point of view, it has been widely recognized that
the absorption and dispersion of GWs in the Universe (filled with a
perfect fluid) should be neglected \citep{Ehlers1996}. However, some
ideas invoked in the context of dark matter (DM) can challenge this
point of view. For instance, many authors have investigated the
possibility of gravitational waves disappearing into dark sector
\citep{Foot16}, i.e., the so-called U(1)$_D$ charged dark matter
with the generation of dark cosmological magnetic fields and
efficient graviton-dark photon conversion \citep{Masaki18}. On the
other hand, if the Universe contains some non-ideal fluids, the
fluid energy-momentum tensor might contain a non-zero shear
viscosity term $\eta$ \citep{Hawking1966}. Within this approach
viscosity provides a damping rate $\beta \equiv 16\pi G\eta$ with
which GWs would be dissipated by matter
\citep{Madore1973,Prasanna1999}. Although the collisionless cold DM
paradigm can successfully account for the observations of large
scale structure \citep{Bahcall1999}, recent progress made in N-body
simulations concerning the small-scale structures has demonstrated
its strong conflict with observations on dwarfs~\citep{Oh2011}, low
surface brightness (LSB) galaxies \citep{Kuzio2008} and clusters
\citep{Newman2013}. In particular, recent works of
\citet{Graziani2020, Marassi2019} should be mentioned in this
context, since they provided the properties (SFR) and evolution
(metallicity evolution) of dwarf galaxies different from large-scale
simulations. More importantly, if DM can be treated as a non-ideal
fluid, the DM self-interaction (SI) can generate the cosmological
shear viscosity \citep{Goswami2017,Atreya2017} and explain the
small-scale structure problems of the Universe \citep{Spergel2000}.
It was recently found that dark matter self-interactions may also
serve as a negative pressure source, which constitutes another
possible mechanism to explain the accelerated expansion of the
Universe \citep{Zimdahl01,Cao11}. In this Letter, we propose that
the combination of unlensed and lensed GW signals would enable
quantitative measurements of GW absorption and dissipation at much
higher redshifts ($z\sim5$), which creates a valuable opportunity to
test the propagation of GWs through the cosmic dark matter.

\section{Method and simulated data}

Up to now, LIGO and Virgo Collaborations have released the
measurements of $D_{L,obs}$ for eleven GW sources. However, it
should be stressed that the luminosity distances $D_{L,obs}$ were
derived from the observed strain $h(t)$ and frequency evolution
$f(t)$ under the assumption that the Universe is filled only with
perfect fluids. DM self-interaction manifesting as effective shear
viscosity of DM component would modify this inference. Namely, as
demonstrated in \citet{Goswami2017,Lu18}, the strain of GW signals
at the co-moving distance of $D$, which then propagate through the
viscous fluid would be modified to
\begin{eqnarray}\label{Ho}
h_{\alpha,visc}= h_{\alpha} e^{-\beta D/2},
\end{eqnarray}
where $\alpha$ index denotes $(+,\times)$ polarizations of the GW.
Considering the inverse relation between the amplitude of GW
waveform and the luminosity distance ($D_L$), let us define the
effective luminosity distance inferred from the GW signal
propagating through the viscous fluid
\begin{eqnarray}\label{DLeff}
D_{L,eff}(z, \beta) = D_L(z) e^{\beta D(z)/2}.
\end{eqnarray}
Once we have independent measurements of the luminosity distance
$D_L(z)$ of the source, for example from the EM counterpart, one can
use the $D_{L,eff}(z)$ derived from the GWs to estimate the damping
rate $\beta$ and thus the shear viscosity $\eta$ of DM fluid. Such
strategy is however hard to realize. We propose another approach by
using strongly lensed GW accompanied by EM signal to test the DM
damping effect on unlensed GW signals. The main advantage of GW
strong lensing over strongly lensed quasars and SNe Ia
\citep{Cao15,Cao18}, is that the time delay between multiple images
can be determined with unprecedented accuracy \citep{Liaokai2017}.

In the general framework of strong gravitational lensing, for a
coalescing compact binary system located at the position of
$\boldsymbol{\beta}$, the time delay between its detected GW signals
$\boldsymbol{\theta}_i$ and $\boldsymbol{\theta}_j$ can be written
as \citep{Treu10}
\begin{equation}
\Delta t_{i,j} = \frac{D_{\mathrm{\Delta
t}}(1+z_{\mathrm{l}})}{c}\Delta \phi_{i,j}, \label{relation}
\end{equation}
where $\Delta\phi_{i,j}$ is the Fermat potential difference
determined by the source position and two-dimensional lensing
potential $\psi$. Here the time-delay distance is defined as
\begin{equation}
D_{\mathrm{\Delta t}}(z_l, z_s) \equiv\frac{D_{A}(z_l)
D_{A}(z_s)}{D_{A}(z_l, z_s)},
\end{equation}
a combination of three angular diameter distances between the
observer ($z=0$), the lens ($z=z_l$) and the source ($z=z_s$). From
Eq.~(\ref{relation}) one can see that $D_{\mathrm{\Delta t}}$ can be
estimated from time delays, provided one is able to reconstruct the
Fermat potential. Such reconstruction is possible if optical images
of the lensing systems are available. Because the speed of GWs has
already been confirmed to be equal to the speed of light with
fractional accuracy of $10^{-16}$, time delays derived from GW are
not affected by DM damping effect. Therefore the measured
$D_{\mathrm{\Delta t}}^{lensed}(z_l,z_s)$ can be tested against
$D_{\mathrm{\Delta t}}^{unlensed}(z_1,z_2; \beta)$ derived from
effective luminosity distances of GW signals originating at
redshifts close to $z_l,z_s$. More precisely, the angular diameter
distance and the luminosity distance are related as
$D_A(z)=D_L(z)/(1+z)^2$. Moreover, in a flat universe one has
$(1+z_s) D_{A}(z_l,z_s) = (1+z_s) D_{A}(z_s) - (1+z_l) D_{A}(z_l)$,
based on which one could derive the time delay distance as
\begin{equation} \label{TDist}
D_{\mathrm{\Delta t}} = \frac{D_{L}(z_l)D_{L}(z_s)}{(1+z_l)^2
D_{L}(z_s) - (1+z_s) (1+z_l) D_{L}(z_l)}
\end{equation}
The Eq.~(\ref{TDist}) merely shows how the time delay distance could
be calculated in terms of luminosity distances. The true, unaffected
value of $D_{\mathrm{\Delta t}}$ (by DM damping effect on the
waveforms) would be obtained from accurate time delays of lensed GW
signals (note that no template fitting is needed for this purpose),
combined with Fermat potential reconstruction following dedicated
examination of the lensing system in the optical. However, the
statistics of unlensed GW events would be much richer than lensed
ones and one would be able to find the unlensed events matched to
$z_s$ and $z_l$ of the GW strong lensing system. In this way one
would be able to reconstruct the time delay distance of lensing
system using measured luminosity distances of unlensed standard
sirens, i.e. the effective luminosity distances which are affected
by viscous DM damping (through Eq.~(\ref{DLeff})).

In this work, we focus on GW signals from binary neutron stars (BNS)
covering the redshift range of $z\sim5$, which could be detected in
the future using the third-generation ground-based technology of the
Einstein Telescope (ET) \citep{Taylor12} and the second-generation
technology of space-borne Big Bang Observer (BBO) \citep{Cutler09}.
It should be mentioned that in the meantime, another space-borne
interferometer, DECi-hertz Interferometer Gravitational wave
Observatory (DECIGO) has also been proposed \citep{Kawamura2011}. It
will be sensitive to GWs at mHz to 100 Hz and capable of observing
GWs from the binaries which would also be targets of ground-based
detectors, such as LIGO/Virgo and the ET \citep{Kawamura2019}.
Different strategies should be applied to derive the redshift of GW
sources: one is identifying electromagnetic counterparts like short
$\gamma$ ray burst (SGRB) and the other is uniquely determining the
host galaxy in the optical (by LSST) and coordinated
optical/infrared campaign, benefit from BBO's unprecedented angular
resolution. We perform Monte Carlo simulations to create the
unlensed and lensed samples. For the unlensed GW events, the
observations of these standard sirens are simulated as follows: (1)
Instead of the specific fitting form of BNS merger rate extensively
used in the literature \citep{Cai15,Qi19}, we apply the
redshift-dependent merger rate of double compact objects, based on
the conservative SFR function from \citep{Dominik13}, to
characterize the probability distribution function of $10^4$ GW
sources in the redshift range of $0<z<5$. The review of cosmic
history of star formation can be found in \citet{Madau2014}, where
the reliable SFR estimates from the observations are provided. In
this analysis we use, however the results of \citet{Dominik13}
obtained with the population synthesis code {\tt StarTrack}, for the
sake of consistency with the strong lensing predictions made using
this model.

(2) The mass of neutron stars in the BNS systems, whose position is
randomly sampled in the interval of $\theta\in[0,\pi]$, will be
sampled from uniform distribution of $m\in[1,2] M_\odot$. Such
methodology follows the recent analysis of both unlensed and lensed
BNS events in the framework of Einstein telescope
\citep{Piorkowska13,Cai15}. (3) Different error strategies will be
implemented to our simulation of luminosity distance $D_{L,eff}(z)$
in ground-based Einstein Telescope (ET) and space-based Big Bang
Observer (BBO). On the one hand, for the ET which is very sensitive
in the frequency range of $1-10^4$ Hz, the amplitude of GW signals
produced by inspiralling binaries could be derived in the Fourier
space, focusing on the post-Newtonian and stationary phase
approximation: \citep{Zhao11}
\begin{eqnarray}
\mathcal{A}=&~~\frac{1}{D_{L,eff}(z)}\sqrt{F_+^2(1+\cos^2(\iota))^2+4F_\times^2\cos^2(\iota)}\nonumber\\
            &~~\times \sqrt{5\pi/96}\pi^{-7/6}\mathcal{M}_c^{5/6},
\label{equa:A}
\end{eqnarray}
where $D_{L,eff}(z_s)$ is the effective luminosity distance of the
source defined above (affected by DM damping), $F_{+,\times}$
represent the beam-pattern functions depending on the position of GW
source ($\theta, \phi$) and the polarization angle ($\alpha$), while
$\iota$ denotes the inclination of the binary's orbital plane. Note
that a criterion of $\iota<20^\circ$ is implemented in our
simulations, considering that SGRB is emitted in a narrow cone
\citep{Nakar07}. Focusing on a binary system consists of component
masses $m_1$ and $m_2$, one could define the chirp mass as
$\mathcal{M}_c=M \nu^{3/5}$, with the total mass of $M=m_1+m_2$ and
the symmetric mass ratio of $\nu=m_1 m_2/M^2$. Given the GW
waveform, the combined signal-to-noise ratio (SNR) of three ET
interferometers is given by
$\rho_{net}=\sqrt{\sum\limits_{i=1}^{3}\left\langle
\mathcal{H}^{(i)},\mathcal{H}^{(i)}\right\rangle}$, the square root
of the inner product of the Fourier strain (in this work we take
$\rho_{net}=8$ as the detector's threshold). In order to
characterize the sensitivity of ET on the strain of GW signals, we
turn to \citet{Zhao11} for the explicit expression of the one-side
noise power spectral density (PSD), $S_h(f)$. Now the instrumental
error on the luminosity distance could be estimated at the level of
$\sigma_{D_{L,eff};net}\simeq 2D_{eff}/\rho_{net}$, if the effective
luminosity distance is uncorrelated with other parameters of GW
sources and detectors \citep{Cai15}. For the BBO, we determine the
corresponding distance uncertainties from the results shown in
Fig.~4 of \citet{Cutler09}, which may be reduced to 1\% percent
accuracy due to the fundamentally self-calibrating capability of
space-based detectors at lower frequencies ($10^{-2}-1$ Hz). On the
hand, the second source of systematic error on $D_{L,eff}(z)$ comes
from the weak-lensing effect, which is quantified as
$\sigma_{D_{L,eff};lens}/D_{L,eff}=0.05z$ for ET in the present
analysis \citep{Sathyaprakash10}. For BBO, we use the fitting
formula of \citet{Cutler09} to estimate the weak lensing uncertainty
of the effective distance:
$\sigma_{D_{L,eff};lens}/D_{L,eff}=0.044z$, which is much larger
than the instrumental uncertainties arising from the detectors.

For the lensed GW events, one can expect that a considerable catalog
of strongly-lensed gravitational waves will be detected, within the
observing strategy of future ground-based and space-ground GW
detectors. Focusing on the detection of lensed BNS mergers by ET and
BBO with their EM counterpart determined by Large Synoptic Survey
Telescope (LSST), in this analysis we carry out the simulation
approach as follows \citep{Collett15,Cao19}: (1) The redshift
distribution of the lensing galaxies is characterized by the
modified Schechter function, i.e., the velocity dispersion
distribution function and their values are taken from SDSS Data
Release 5 of elliptical galaxies \citep{Choi07}. Meanwhile, a
spherically symmetric power-law model ($\rho\sim r^{-\gamma}$) is
applied to quantify the mass distribution of each galactic-scale
lens, with the average density slope of SLACS lenses
($\gamma=2.085\pm0.02$) as our fiducial lens model
\citep{Koopmans09}. The final results show that for the simulated
population of realistic lensing systems, the distributions of the
lens redshift ($z_l=0.8$) and the velocity dispersion
($\sigma_{0}=210\pm50$ km/s) are well consistent with those of the
available LSD lenses. (2) In each simulation, we assume that ET will
conservatively detect 10 lensing events (with their EM counterparts)
during its successful operation \citep{Ding15}, while $10^3$ such
events would be possibly registered by BBO from the observations of
$10^6$ BNS systems \citep{Cutler09}. Concerning the sampling
distribution of the lensed BNS population, we adopt the same
approach as in \citet{Ding15} and obtain the sampling distribution
of lensed GWs, from the well-known "rest frame rates" calibrated by
strong lensing effects. (3) In our simulation of lensed GWs, three
ingredients are considered to contribute the uncertainty of the time
delay distance. Firstly, compared with traditional strong lensing
systems which monitor light curves of quasars and SNe Ia, the time
delays measured from lensed GWs can be accurately determined given
their negligible uncertainties due to the short duration of GW
signals from BNS mergers ($\sim$ 0.1s) \citep{Liaokai2017}.
Secondly, benefit from high-resolution imaging of the EM counterpart
(i.e., the host galaxy), the absence of central dazzling active
galactic nuclei (AGN) could decrease the fractional uncertainty of
Fermat potential difference to the level of $\sim 0.5\%$
\citep{HOLI}. Finally, similar to the strong lensing systems in the
EM domain, the lensed GWs are still confronted with the
light-of-sight (LOS) contamination, which will introduce an
additional 1\% uncertainty in the Fermat potential reconstruction
\citep{Cao19}.

\begin{figure}
\includegraphics[scale=0.32]{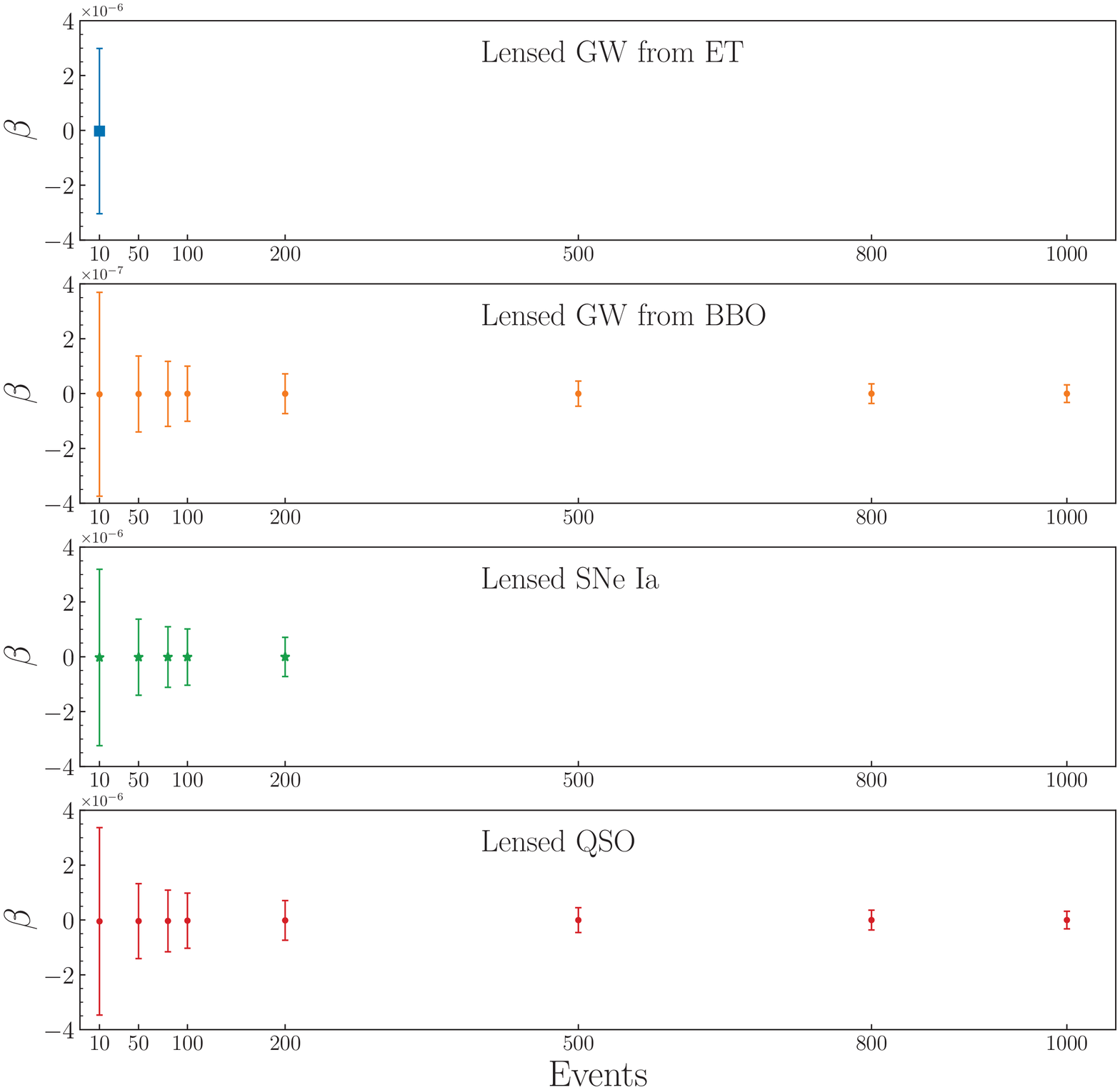}
\caption{Fits on the viscosity of dark matter halos with different
number of strongly lensed sources in both GW (lensed GWs) and EM
domains (lensed SNe Ia and quasars). }\label{fig1}
\end{figure}

\begin{table*}
\centering \setlength{\tabcolsep}{5mm}{
\begin{tabular}{l|ccc}
\hline \hline
Data & $\Delta \beta$ & $\Delta(\sigma_{\chi}/m_{\chi}) (I)$  & $\Delta(\sigma_{\chi}/m_{\chi}) (II)$ \\
GW (Lensed; ET) + GW (Unlensed) & $10^{-6}$ Mpc$^{-1}$ & $10^{-4}$ cm$^2$/g & $10^{-3}$ cm$^2$/g \\
GW (Lensed; BBO) + GW (Unlensed) & $10^{-8}$ Mpc$^{-1}$ & $10^{-6}$ cm$^2$/g & $10^{-5}$ cm$^2$/g \\

\hline

QSO (Lensed; LSST) + GW (Unlensed) & $10^{-7}$ Mpc$^{-1}$ & $10^{-5}$ cm$^2$/g & $10^{-4}$ cm$^2$/g \\
SNe Ia (Lensed; LSST) + GW (Unlensed) & $10^{-6}$ Mpc$^{-1}$ & $10^{-4}$ cm$^2$/g &  $10^{-3}$ cm$^2$/g \\

\hline \hline
\end{tabular}}
\caption{Summary of the constraints obtained from different
observations. Type I and II respectively correspond to the two cases
of self-interacting dark matter in galaxies and galaxy clusters. }
\label{table1}
\end{table*}

\section{Implementation and results}

Now we will illustrate what kind of results could be derived from
our method, based on the collisionless cold DM ($\eta=0$) and
$\Lambda$CDM cosmology from $Planck+WMAP+highL+BAO$ \citep{Ade16}.
The final results are shown in Table 1, based on the simulation
process repeated $10^3$ times for each lensed and unlensed GW data
set. More specifically, in our analysis the $\beta$ parameter is
determined in the framework of Markov Chain Monte Carlo (MCMC)
minimizations, with the corresponding $\chi^2$ defined as
\begin{equation}
\chi^2 = \sum_{i=1}^{i} \frac{\left( D_{\Delta
t,i}^{lens}(z_{l,i},z_{s,i})-D_{\Delta
t,i}^{unlens}(z_{l,i},z_{s,i}; \beta)
\right)^2}{\sigma_{i,lens}^2+\sigma_{i,unlens}^2}.
\end{equation}
Note that $D_{\Delta t,i}^{lens}$ denotes the time delay distance
for the $ith$ lensed GW system, with the total uncertainty of
$\sigma_{i,lens}$. Its corresponding counterpart from the unlensed
GW observations is represented by $D_{\Delta t,i}^{unlens}$, with
the total uncertainty of $\sigma_{i,unlens}$ (statistical and
systematic uncertainties).

For the ET with 10 strongly lensed GW events acting as
multi-messengers, we would obtain a constraint on the damping rate
with uncertainty $\Delta \beta=10^{-6}\,{\rm Mpc}^{-1}$. The
resulting constraint on the $\beta$ parameter become $\Delta
\beta=10^{-8}\,{\rm Mpc}^{-1}$ for the BBO, with larger number of
strongly lensed GW events detected in the second-generation
technology. Such stringent test of the damping of GWs in a viscous
universe can also be confidently used to perceive dark matter
self-interactions. Now we can attempt to put constraints on the dark
matter scatter cross-section ($\sigma_{\chi}/m_{\chi}$) from the
gravitational wave damping rate. In the framework of
self-interacting dark matter scenario \citep{Tulin2017}, in which DM
self-scattering is efficient for hydrodynamic description to be
valid and assigning a Maxwellian distribution for DM particles, the
DM SI cross section $ \sigma_\chi $ can be related to the shear
viscosity $\eta$ as $\eta =\frac{1.18\,m_{\chi}\left \langle v
\right \rangle}{3 \, \sigma_{\chi}}$, where $m_{\chi}$ and $\langle
v \rangle$ are the mass and average velocity of DM particles,
respectively \citep{Atreya2017}. It is straightforward to obtain the
above relation in terms of the GW damping rate as
\begin{eqnarray}
\frac{\sigma_{\chi}}{m_{\chi}}=\frac{6.3\pi G\left \langle v \right
\rangle}{\beta}.
\end{eqnarray}

The main question to be addressed is: \textit{Is such precision high
enough to detect possible DM shear viscosity or self-interactions in
gravitational wave domain?} For our analysis, we have considered two
scenarios: one with the DM SI cross sections and average velocity
deduced from the fittings to dwarf galaxies, and low surface
brightness (LSB) galaxies; and the other with the DM SI cross
sections and average velocity deduced from the fittings to galaxy
clusters \citep{Kaplinghat2016}. Let us first consider the SI model
deduced from the fittings to dwarf galaxies and LSB galaxies (the
velocity-dependent cross section is $\sigma_{\chi}/m_{\chi} \sim 1$
cm$^2$/g on galaxy scales). When considering the intrinsic
uncertainty in the mean collision velocity $\Delta\langle v
\rangle\sim10^2$ km/s, 10 lensed GWs detected by ET will be able to
detect (if any) galactic-scale DM SI cross section at the precision
of $\Delta(\sigma_{\chi}/m_{\chi}) \sim 10^{-4}$ cm$^2$/g. Focusing
on the SI model corresponding to dark matter in galaxy clusters we
have $\Delta(\sigma_{\chi}/m_{\chi}) \sim 10^{-3}$ cm$^2$/g,
considering the intrinsic uncertainty in the mean collision velocity
$\Delta\langle v \rangle\sim10^3$ km/s. Therefore, ET will also
succeed in detecting cluster-scale DM SI at very high confidence
level, since the velocity-dependent cross section is $\sigma/m\sim
10^{-1}$ cm$^2$/g on cluster scales, consistent with the upper
limits from merging clusters \citep{Kaplinghat2016}. In the
framework of BBO, the second-generation space-based GW detector, it
is expected to detect galaxy-scale DM SI with the precision reaching
at even $\Delta(\sigma_{\chi}/m_{\chi}) \sim 10^{-6}$ cm$^2$/g.
Significant improvements would also be obtained for the
cluster-scale case: $\Delta(\sigma_{\chi}/m_{\chi}) \sim 10^{-5}$
cm$^2$/g. Therefore, we expect such stringent DM SI constraints from
the GW damping can probe the DM self-scattering solution to the
small-scale structure problems \citep{Kaplinghat2016}.

There are a few other ways in which our technique might be
implemented. In our approach the traditional strong lensing systems
in the EM domain, i.e., strongly lensed quasars and SNe Ia with
well-measured time delays \citep{Goobar17} could also be used
instead of lensed GWs. More importantly, the upcoming LSST
accompanied by a long baseline multi-epoch and 10 year $z$-band
observational campaign, will enable the discovery of $10^3$
quasar-elliptical galaxy systems and 200 SNe Ia-elliptical galaxy
systems with well-measured time-delays \citep{Oguri10}. Following
the recent analysis by the TDC \citep{Dobler15}, the $\Delta t$
measurements may achieve the precision of $\sim$3\% and 1\%
including systematics. We remark here that the well-characterized
spectral sequences of lensed SNe Ia will significantly contribute to
the reduction of relative uncertainty from time delay measurements,
which has been extensively discussed in the literature
\citep{Nugent02,Pereira13}. Meanwhile, the average relative
uncertainty of lens modeling with high-resolution imaging will reach
to the level of $\sim$3\%, due to the inevitable effects of dazzling
AGNs in the source center \citep{Suyu13}. Actually, as can be
clearly seen from the results shown in Table I, such combination of
strongly lensed SNe Ia or quasars in EM domain will also results in
stringent constraints on the viscosity of dark matter halos, similar
to the precision obtained from lensed GWs. For a better comparison,
Fig.~1 shows the $\beta$ parameter assessment as a function of
sample sizes.

Finally, we need to comment about the technical barriers that could
potentially affect precise measurements of DM self-scattering. In
our approach based on lensed GW+EM events, a high temporal
resolution EM detector is also necessary to monitor the multiple
images of the EM event. Qualitatively, one can expect that such
issue will be suitably addressed in the future analyses. As was
extensively discussed concerning the observational schemes of many
on-going and planned GW detectors in the near future
\citep{Taylor12}, the source location uncertainties of ET will be
much smaller than those of the joint detections by Advanced LIGO and
Virgo. Spaced based detectors are much more competitive in this
respect than ground-based ones. Therefore, considering the strong
correlation between gravitational wave and its electromagnetic
counterparts (SGRB, etc.), it is surely possible to detect the
strongly lensed GW and EM events simultaneously, based on the
combination of future GW detectors (both ground-based and
space-based detectors) and their follow-up facilities in the EM
domain. Summarizing, given the expected wealth of gravitational
lensing data in GW and EM domains, we may hopefully obtain precise
measurements of the viscosity of dark matter in the early Universe
($z\sim5$), which could shed lights on our understanding of dark
matter self-interactions at both galaxy and cluster scales
\citep{Cao11}.

\section*{Acknowledgments}

This work was supported by National Key R\&D Program of China No.
2017YFA0402600; the National Natural Science Foundation of China
under Grants Nos. 12021003, 11690023, 11633001, and 11373014;
Beijing Talents Fund of Organization Department of Beijing Municipal
Committee of the CPC; the Strategic Priority Research Program of the
Chinese Academy of Sciences, Grant No. XDB23000000; and the
Interdiscipline Research Funds of Beijing Normal University.

\section*{Data Availability Statements}

The data underlying this article will be shared on reasonable
request to the corresponding author.

%%%%%%%%%%%%%%%%%%%%%%%%%%%%%%%%%%%%%%%%%%%%%%%%%%%%%%%%%%%%%%%%%%%%%%%%%%%%%%
%%%%%%%%%%%%%%%%%%%%%%%%%%%%%%%%%%%%%%%%%%%%%%%%%%%%%%%%%%%%%%%%%%%%%%%%%%%%%%

\label{lastpage}

\end{document}